\documentstyle[proceedings]{crckapb}

\let\sec=\section
\let\ssec=\subsection

\def\erf{{\rm erf}}
\def\japref{\parskip=0pt\par\noindent\hangindent\parindent
    \parskip =2ex plus .5ex minus .1ex}

\def\gs{\mathrel{\lower0.6ex\hbox{$\buildrel {\textstyle >}
 \over {\scriptstyle \sim}$}}}
\def\ls{\mathrel{\lower0.6ex\hbox{$\buildrel {\textstyle <}
 \over {\scriptstyle \sim}$}}}
\newcount\equationnum
\global\equationnum=0
\def\bookdisp#1$${\leftline{\hfill{$\displaystyle#1$}
    \global\advance\equationnum by 1
    \hfill (\the\equationnum )}$$}
\everydisplay{\bookdisp}
\def\ss{\rm\scriptscriptstyle}

\def\kms{{\,\rm km\,s^{-1}}}

\def\hompc{{\,h\,\rm Mpc^{-1}}}
\def\mpcoh{{\,h^{-1}\,\rm Mpc}}
\def\japitem#1{\medskip\noindent\rlap{#1}\hglue 3em\hangindent 3em}

\def\key#1{#1}

\def\vec#1{
\setbox0=\hbox{$#1$}           
   \dimen0=\wd0  
   \wd0=0.7\wd0
   \rlap{\hbox to \dimen0{\hfil\underline{\phantom{\box0}}\hglue 1pt\hfil}}    
#1
}

\def\year#1{ (#1)}

\def\mnras{Mon. Not. R. Astr. Soc.}
\def\mn{Mon. Not. R. Astr. Soc.}

\def\apj{Astrophys. J.}
\def\apjs{Astrophys. J. Suppl.}

\def\prl{Phys. Rev. Lett.}

\input epsf

\def\japfig#1#2#3#4#5#6{
\begin{figure*}
\centering\mbox{\epsfxsize=0.9\hsize\epsfbox[#1 #2 #3 #4]{#5}}
\caption[]{#6}
\end{figure*}
}

\def\m@th{\mathsurround=0pt }
\def\eqalign#1{\null\,\vcenter{\openup1\jot \m@th
 \ialign{\strut\hfil$\displaystyle{##}$&$\displaystyle{{}##}$\hfil
 \crcr#1\crcr}}\,}

\def\topinsert{\begin{figure}}
\def\endinsert{\end{figure}}

\begin{opening}
\title{CLUSTERING OF MASS AND GALAXIES}
\author{J.A. PEACOCK}
\institute{Institute for Astronomy,
University of Edinburgh\\
Royal Observatory, Edinburgh EH9 3HJ, UK}
\end{opening}
\begin{document}

\vglue -7.5truecm
\centerline{
Invited lectures at the NATO ASI {\it Structure Formation in the Universe\/}}
\smallskip
\centerline{Cambridge, August 1999}
\vglue 6.5truecm
\noindent

\begin{abstract}
These lectures cover various aspects of the
statistical description of cosmological
density fields. Observationally, this consists
of the point process defined by galaxies, and the
challenge is to relate this to the continuous
density field generated by gravitational instability
in dark matter. The main topics discussed are
(1) nonlinear structure in CDM models;
(2) statistical measures of clustering;
(3) redshift-space distortions;
(4) small-scale clustering and bias.
The overall message is optimistic, in that
simple assumptions for where galaxies should
form in the mass density field allow one
to understand the systematic differences
between galaxy data and the predictions of
CDM models.
\end{abstract}

\sec{Preamble}

The subject of large-scale structure is in
a period of very rapid development.
For many years, this term would have meant only
one thing: the distribution of galaxies.
However, we are increasingly able to probe the
primordial fluctuations through the CMB, so that
the problem of galaxy formation and clustering is
now only one aspect of the general picture of
structure formation.
The rationale for studying the large-scale
distribution of galaxies is therefore altering.
Ten years ago, we were happy to produce samples
based on a rather sparse random sampling of the
galaxy distribution, with the main aim of
tying down statistics such as the large-scale
power spectrum of number-density fluctuations.
A major goal of the subject remains the measurement
of the fluctuation spectrum for wavelengths
$\gs 100$~Mpc, and the demonstration that this
agrees in shape with what can be inferred from the
CMB. Nevertheless, we are now increasingly interested
in studying the pattern of galaxies with the highest
possible fidelity --  demanding deep, fully-sampled
surveys of the local universe. Such studies will
tell us much about the processes by which galaxies formed
and evolved within the distribution of dark
matter. The aim of these lectures is therefore to
look both backwards and forwards: reviewing the
foundations of the subject and looking forward to the
future issues.

\sec{The CDM family album}

\ssec{The linear spectrum}

The basic picture of inflationary models (but also
of cosmology before inflation) is of a
primordial power-law spectrum, written dimensionlessly
as the logarithmic contribution to the fractional
density variance, $\sigma^2$:
$$
\Delta^2(k)={d\sigma^2\over d\ln k} \propto k^{3+n},
$$
where $n$ stands for $n_{\ss S}$ hereafter.
This undergoes linear growth
$$
\delta_k(a) = \delta_k(a_0)\; \left[{D(a)\over D(a_0)}\right] \; T_k,
$$
where the linear growth law is
$$
D(a)=a\, g[\Omega(a)]
$$
in the matter era,
and the growth suppression for low $\Omega$ is
$$
\eqalign{
g(\Omega) &\simeq \Omega^{0.65}\ {\rm (open)} \cr
&\simeq \Omega^{0.23}\ {\rm (flat)} \cr
}
$$
The transfer function $T_k$ depends on the dark-matter
content as shown in figure 1.

\japfig{31}{183}{499}{579}{japfig1.eps}
{Transfer functions for various dark-matter models.
The scaling with $\Omega h^2$ is exact only for the
zero-baryon models; the baryon results are scaled from
the particular case $\Omega_{\ss B}=1$, $h=1/2$. 
}

Note the baryonic oscillations in figure 1; these
can be significant even in CDM-dominated models
when working with high-precision data.
Eisenstein \& Hu (1998) are to be congratulated for
their impressive persistence in finding an accurate
fitting formula that describes these wiggles.
This is invaluable for carrying out a search of
a large parameter space.

\japfig{0}{0}{519}{582}{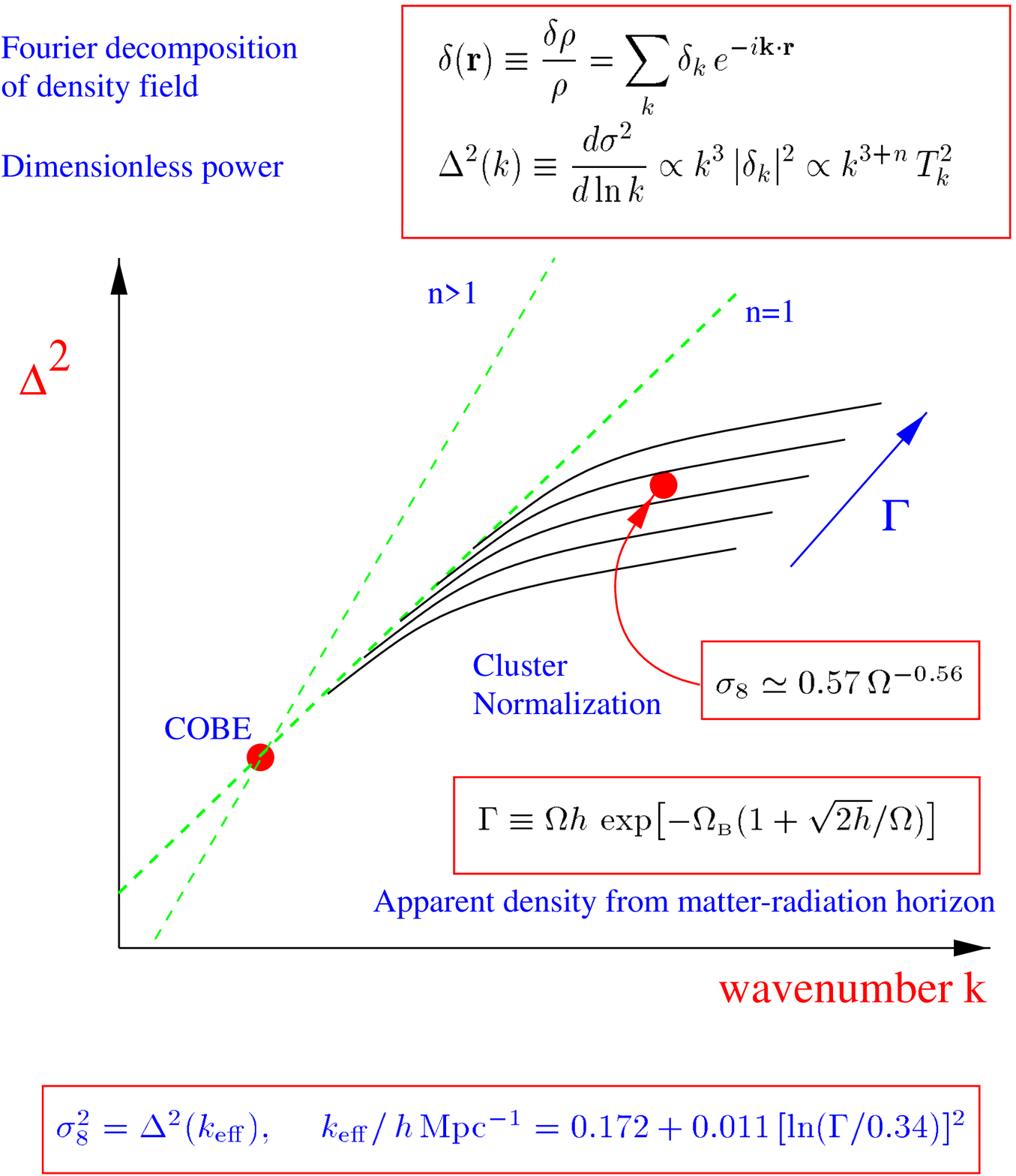}
{This figure
illustrates how the primordial power spectrum is
modified as a function of density in a CDM
model. For a given tilt, it is always
possible to choose a density that satisfies both the
COBE and cluster normalizations.}

\japfig{36}{192}{482}{578}{japfig3.eps}
{For 10\% baryons, the value of $n$ needed
to reconcile COBE and the cluster normalization
in CDM models.}

The state of the linear-theory spectrum after these
modifications is illustrated in figure 2.
The primordial power-law spectrum is reduced
at large $k$, by an amount that depends on
both the quantity of dark matter and its
nature. Generally the bend in the spectrum
occurs near $1/k$ of order the horizon size
at matter-radiation equality, $\propto 
(\Omega h^2)^{-1}$. For a pure CDM universe,
with scale-invariant initial fluctuations
($n=1$), the observed spectrum depends only on two
parameters. One is the shape $\Gamma = \Omega h$,
and the other is a normalization. On the
shape front, a government health warning is needed,
as follows. It has been quite common to take
$\Gamma$-based fits to observations as indicating
a {\it measurement\/}  of $\Omega h$, but there are
three reasons why this may give incorrect answers:

(1) The dark matter may not be CDM. An admixture of
HDM will damp the spectrum more, mimicking a
lower CDM density.

(2) Even in a CDM-dominated universe, baryons can
have a significant effect, making $\Gamma$ lower
than $\Omega h$. An approximate formula for this
is given in figure 2 (Peacock \& Dodds 1994;
Sugiyama 1995).

(3) The strongest (and most-ignored) effect is
tilt: if $n\ne 1$, then even in a pure CDM universe
a $\Gamma$-model fit to the spectrum will give a
badly incorrect estimate of the density
(the change in $\Omega h$ is roughly $0.3(n-1)$;
Peacock \& Dodds 1994).

\ssec{Normalization}

The other parameter is the normalization.
This can be set at a number of points.
The COBE normalization comes from large angle
CMB anisotropies, and is sensitive to the
power spectrum at $k\simeq 10^{-3}\hompc$.
The alternative is to set the normalization
near the quasilinear scale, using the abundance of
rich clusters. Many authors have tried this
calculation, and there is good agreement on the
answer:
$$
\sigma_8 \simeq (0.5 - 0.6) \, \Omega_m^{-0.6}.
$$
(White, Efstathiou \& Frenk 1993; Eke et al. 1996; Viana \& Liddle 1996).
In many ways, this is the most sensible normalization
to use for LSS studies, since it does not rely
on an extrapolation from larger scales.

Within the CDM model, it is always possible to satisfy
both these normalization constraints, by appropriate
choice of $\Gamma$ and $n$. This is illustrated in
figure 3. Note that vacuum energy affects the answer;
for reasonable values of $h$ and reasonable
baryon content, flat models require $\Omega_m\simeq 0.3$, whereas
open models require $\Omega_m\simeq 0.5$.

\ssec{The nonlinear spectrum}

On smaller scales ($k\gs 0.1$), nonlinear effects become important.
These are relatively well understood so far as they affect the
power spectrum of the mass (e.g. 
Hamilton et al. 1991; Jain, Mo \& White 1995;
Peacock \& Dodds 1996). 
Based on a fitting formula for the similarity solution
governing the evolution of scale-free initial conditions,
it is possible to predict the evolved spectrum in CDM
universes to a few per cent precision (e.g. Jenkins et al. 1998).

These methods can cope with most smoothly-varying power
spectra, but they break down for models with a large
baryon content. Figure 1 shows that rather large oscillatory
features would be expected if the universe was baryon
dominated. The lack of observational evidence for 
such features is one reason
for believing that the universe might be dominated
by collisionless nonbaryonic matter (consistent with
primordial nucleosynthesis if $\Omega_m\gs 0.1$).

\japfig{0}{25}{487}{544}{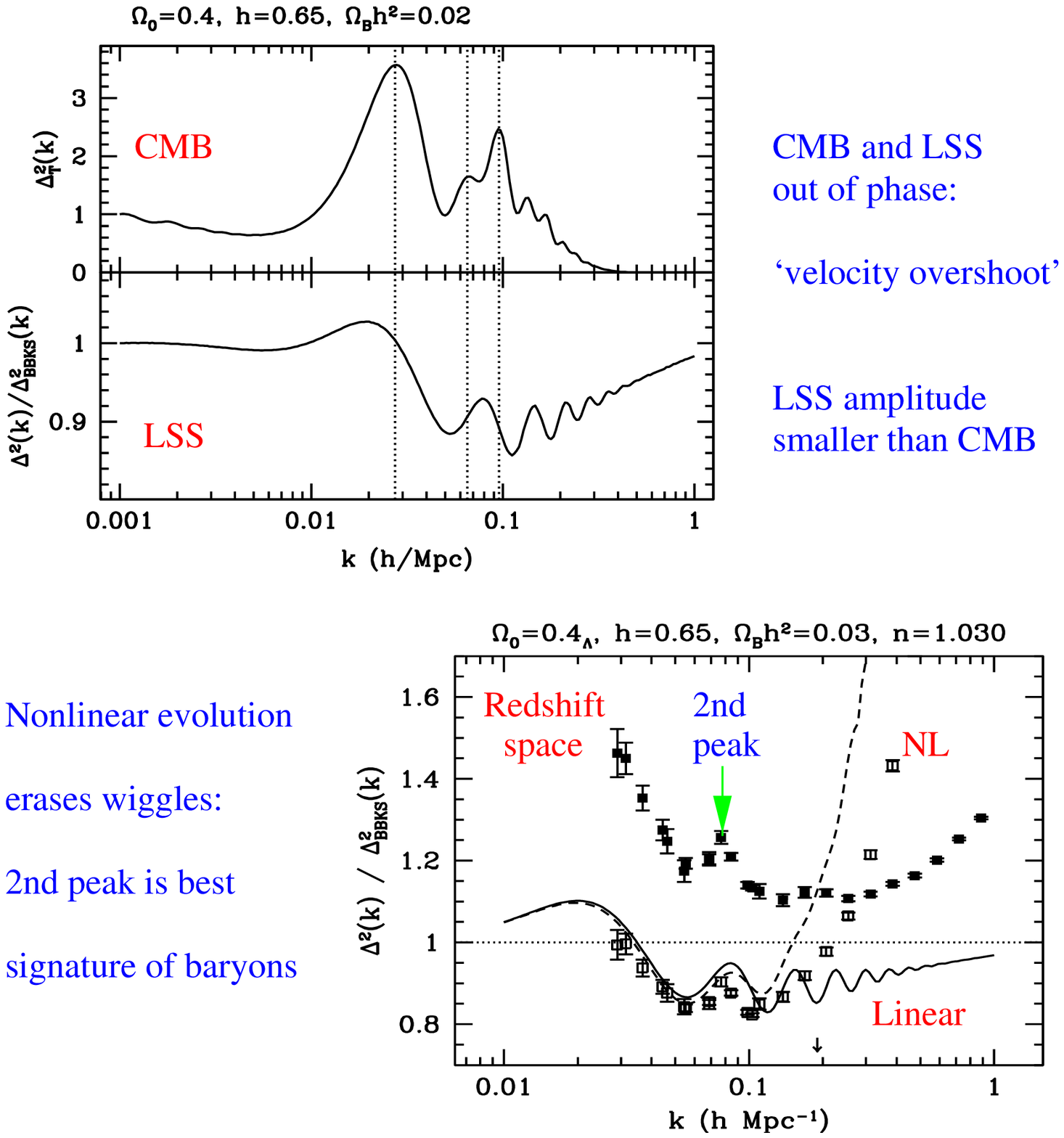}
{Baryonic fluctuations in the spectrum can
become significant for high-precision measurements.
Although such features are much less important in the
density spectrum than in the CMB (first panel), the
order 10\% modulation of the power is potentially
detectable. However, nonlinear evolution has the
effect of damping all beyond the second
peak. This second feature is relatively
narrow, and can serve as a clear proof of the past
existence of oscillations in the baryon-photon fluid
(Meiksin, White \& Peacock 1999).}

Nevertheless, 
baryonic fluctuations in the spectrum can
become significant for high-precision measurements.
Figure 4 shows that order 10\% modulation of the
power may be expected in realistic baryonic
models (Eisenstein \& Hu 1998; Goldberg \& Strauss 1998).
Most of these features are however removed
by nonlinear evolution.
The highest-$k$ feature to survive is usually
the second peak, which almost always lies
near $k=0.05\,{\rm Mpc}^{-1}$ (no $h$, for a change).
This feature is relatively
narrow, and can serve as a clear proof of the past
existence of baryonic oscillations in forming the 
mass distribution
(Meiksin, White \& Peacock 1999).
However, figure 4 emphasizes that the easiest way
of detecting the presence of baryons is likely to
be through the CMB spectrum. The oscillations
have a much larger `visibility' there, because
the small-scale CMB anisotropies come directly
from the coupled radiation-baryon fluid, rather
than the small-scale dark matter perturbations.

\sec{Statistics}

Statistical measures of the cosmological density field
relate to properties of
the dimensionless \key{density perturbation field}
$$
\delta ({\bf x}) \equiv {\rho({\bf x})-\langle\rho\rangle \over
\langle\rho\rangle},
$$
although $\delta$ need not be assumed to be small. 

\ssec{Correlation functions}
The simplest measure is the \key{autocorrelation function}
of the density perturbation
$$
\xi_{\ss A}({\bf r}) \equiv \left\langle \delta({\bf x})\delta({\bf x+ r})\right\rangle,
$$
This is a straightforward statistical measure that can
also be computed for the dark-matter distribution in
$N$-body simulations.
Formally, the averaging operator here is an ensemble
average, but one generally appeals to the ergodic nature
of the density field to replace this with a volume average.

However, galaxies are a point process, so what astronomers can
measure in practice is the
\key{two-point correlation function}, which gives the excess
probability for finding a neighbour a distance $r$
from a given galaxy. By regarding this as the
probability of finding a pair with one object in each of the volume
elements $dV_1$ and $dV_2$, 
$$
dP=\rho_0^2\, [1+\xi_2(r)]\, dV_1\, dV_2.
$$
Is it true that $\xi_{\ss A}(r)=\xi_2(r)?$ Life would
certainly be simple if so, and much work on large-scale
structure has implicitly assumed the
\key{Poisson clustering hypothesis}, in which galaxies
are assumed to be sampled at random from some continuous
underlying density field. Many of the puzzles in the field
can however be traced to the fact that this hypothesis
is probably false, as discussed below.

A related quantity is the \key{cross-correlation function}.
Here, one considers two different classes of
object (a and b, say), and the cross-correlation function
$\xi_{ab}$ is defined as the (symmetric) probability
of finding a pair in which $dV_1$ is occupied by an object
from the first catalogue and $dV_2$ by one from the second.
Both cross- and auto-correlation functions are readily extended
to higher orders and considerations of $n$-tuples of points
in a given geometry.

\ssec{Fourier space}
For the Fourier counterpart of this analysis, we assume
that the field is periodic within some box of side $L$,
and expand as a Fourier series:
$$
\delta({\bf x}) = \sum \delta_{\bf k} e^{-i{\bf k\cdot x}}.
$$
For a real field, $\delta_{\bf k}(-{\bf k})=\delta^*_{\bf k}({\bf k})$.
Using this definition in the correlation function, most
cross terms integrate to zero through the periodic boundary conditions, giving
$$
\xi({\bf  r})=
{V \over (2\pi)^3}\int|\delta_{\bf k}|^2 e^{-i{\bf  k\cdot r}} d^3 k.
$$
In short, the correlation function is the Fourier transform of
the \key{power spectrum}. 

We shall usually express
the power spectrum in dimensionless form, as the variance per $\ln k$
($\Delta^2(k) =d \langle \delta^2 \rangle/d\ln k \propto k^3 P[k]$):
$$
\Delta^2(k)\equiv {V\over (2\pi)^3} \, 4\pi k^3\, P(k)
={2\over \pi}k^3\int_0^\infty\xi(r)\,
{\sin kr\over kr}\, r^2\, dr.
$$
This gives a more easily visualizable meaning to the power
spectrum than does the quantity $V P(k)$, which has
dimensions of volume: $\Delta^2(k)=1$ means that there
are order-unity density fluctuations from modes
in the logarithmic bin around wavenumber $k$.
$\Delta^2(k)$ is therefore the natural choice for
a Fourier-space counterpart to the dimensionless quantity $\xi(r)$.

In the days before inflation, the primordial power
spectrum was chosen by hand, and the minimal
assumption was a featureless power law:
$$
\left\langle|\delta_k|^2\right\rangle \equiv P(k) \propto k^n
$$
The index $n$ governs the balance between large-
and small-scale power.
Similarly, a power-law spectrum implies a power-law
correlation function.
If $\xi(r)=(r/r_0)^{-\gamma}$, with $\gamma=n+3$,
the corresponding 3D power spectrum is
$$
\Delta^2(k)={2\over\pi}\,(kr_0)^{\gamma}\, \Gamma(2-\gamma) \,
   \sin {(2-\gamma)\pi\over 2}
\equiv \beta (kr_0)^\gamma
$$
($=0.903 (kr_0)^{1.8}$ if $\gamma=1.8$).
This expression is only valid for $n<0$ ($\gamma<3$);
for larger values of $n$, $\xi$ must become
negative at large $r$ (because $P(0)$ must vanish,
implying $\int_0^\infty \xi(r)\, r^2\, dr=0$).
A cutoff in the spectrum at large $k$ is needed
to obtain physically sensible results.

The most interesting value of $n$ is the \key{scale-invariant spectrum}, 
$n=1$, {\it i.e.} $\Delta^2\propto k^4$. To see how the name arises,
consider a perturbation $\delta\Phi$ in the gravitational potential:
$$
\nabla^2\delta\Phi= 4\pi G\rho_0\delta
\quad\Rightarrow\quad \delta\Phi_k = -4\pi G\rho_0\delta_k/k^2.
$$
The two powers of $k$ pulled down by $\nabla^2$ mean
that, if $\Delta^2\propto k^4$ for the power spectrum of 
density fluctuations, then $\Delta^2_\Phi$ is a constant.
Since potential perturbations govern the flatness
of spacetime, this says that the scale-invariant
spectrum corresponds to a metric that is
a \key{fractal}: spacetime has the same degree of
`wrinkliness' on each resolution scale.
The total curvature fluctuations diverge, but only
logarithmically at either extreme of wavelength.

\ssec{Error estimates}

A key question for these statistical measures is how accurate
they are -- i.e. how much does the result for a given
finite sample depart from the ideal statistic averaged over
an infinite universe? Terminology here can be confusing,
in that a distinction is sometimes made between
\key{sampling variance} and \key{cosmic variance}.
The former is to be understood as arising from probing
a given volume only with a finite number of galaxies
(e.g. just the bright ones), so that $\sqrt{N}$
statistics limit our knowledge of the mass distribution
within that region. The second term concerns whether
we have reached a fair sample of the universe, and 
depends on whether there is significant power in
density perturbation modes with wavelengths larger than
the sample depth. Clearly, these two aspects are 
closely related.

The quantitative analysis of these errors is most simply
performed in Fourier space, and was given by 
Feldman, Kaiser \& Peacock (1994). The results can be
understood most simply by comparison with an idealized
complete and uniform survey of a volume $L^3$, with
periodicity scale $L$. For an infinite survey, the arbitrariness of the spatial
origin means that different modes are uncorrelated:
$$
\langle \delta_k({\bf k}_{i})\delta_k^*({\bf k}_{j})\rangle = P(k) \delta_{ij}.
$$
Each mode has an exponential distribution in power (because the complex
coefficients $\delta_k$ are 2D Gaussian-distributed  variables  on the
Argand plane), for which the mean and rms are identical. The fractional
uncertainty in the mean power measured over some $k$-space volume is
then just determined by the number of uncorrelated modes averaged over:
$$
{\delta \bar P \over \bar P} = {1\over N_{\rm modes}^{1/2}}; \quad\quad
N_{\rm modes} = \left({L\over 2\pi}\right)^3\, \int d^3 k.
$$
The only subtlety is that, because the density field is real, modes
at $k$ and $-k$ are perfectly correlated. Thus, if the $k$-space
volume is a shell, the effective number of uncorrelated modes is
only half the above expression.

Analogous results apply for an arbitrary survey selection function.
In the continuum limit, the Kroneker delta in the 
expression for mode correlation would be
replaced a term proportional to a delta-function, $\delta[{\bf k}_{i}-{\bf k}_{j}]$).
Now, multiplying the infinite ideal survey by a survey window, $\rho({\bf r})$,
is equivalent to convolution in
the Fourier domain, with the result that the power per mode is
correlated over $k$-space separations of order $1/D$, where $D$ is the
survey depth.

Given this expression for the fractional power, it is clear that the
precision of the estimate can be manipulated by appropriate
weighting of the data: giving increased weight to the most distant
galaxies increases the effective survey volume, boosting the
number of modes. This sounds too good to be true, and of course
it is: the above expression for the fractional power error
applies to the sum of true clustering power and shot noise. 
The latter arises because we transform a point process. Given a set
of $N$ galaxies, we would estimate Fourier coefficients via
$\delta_k=(1/N) \sum_i \exp(-i{\bf k}\cdot x_i)$. From this, the
expectation power is
$$
\langle|\delta_k|^2 \rangle = P(k) + 1/N.
$$
The existence of an additive discreteness correction is no problem,
but the {\it fluctuations\/} on the shot noise hide the 
signal of interest. Introducing
weights boosts the shot noise, so there is an optimum choice of weight that
minimizes the uncertainty in the power after shot-noise subtraction.
Feldman, Kaiser \& Peacock (1994) showed that this weight is
$$
w = (1+\bar n P)^{-1},
$$
where $\bar n$ is the expected galaxy number density as a function of
position in the survey.

Since the correlation of modes arises from the survey
selection function, it is clear that weighting the data
changes the degree of correlation in $k$ space.
Increasing the weight in low-density
regions increases the effective survey volume, and
so shrinks the $k$-space coherence scale.
However, the coherence scale continues to shrink as distant
regions of the survey are given greater weight, whereas the noise
goes through a minimum. There is 
thus a trade-off between the competing desirable criteria of
high $k$-space resolution and low noise.
Tegmark (1996) shows how weights may be chosen to
implement any given prejudice concerning the relative
importance of these two criteria.
See also Hamilton (1997b,c) for similar arguments.

\ssec{Karhunen-Lo\`eve and all that}

Given these difficulties with correlated
results, it is attractive to seek a method
where the data can be decomposed into a set
of statistics that are completely uncorrelated with each other.
Such a method is provided by the
Karhunen-Lo\`eve formalism. Vogeley \& Szalay (1996)
argued as follows.
Define a column vector of data $\vec{d}$;
this can be quite abstract in nature, and could
be e.g. the numbers of galaxies in a set of cells, or
a set of Fourier components of the transformed galaxy
number counts. Similarly, for CMB studies, $\vec{d}$ could
be $\delta T/T$ in a set of pixels, or spherical-harmonic
coefficients $a_{\ell m}$.
We assume that the mean can be identified and
subtracted off, so that $\langle \vec{d} \rangle =0$ in
ensemble average. The statistical properties of the data are
then described by the covariance matrix
$$
C_{ij} \equiv \langle d_i d_j^* \rangle
$$
(normally the data will be real, but it is
convenient to keep things general and include the complex
conjugate). 

Suppose we seek to expand the datavector in
terms of a set of new orthonormal vectors:
$$
{\vec{d}} = \sum_i a_i {\vec{\psi}}_i; \quad\quad
{\vec{\psi}}^*_i \cdot {\vec{\psi}}_j = \delta_{ij}.
$$
The expansion coefficients are extracted in the usual way:
$\smash{a_j = \vec{d} \cdot \vec{\psi}_j^*}$.
Now require that these coefficients be statistically
uncorrelated,
$\langle a_i a_j^* \rangle = \lambda_i \delta_{ij}$
(no sum on $i$). This gives
$$
\vec{\psi}_i^* \cdot \langle \vec{d} \, \vec{d}^* \rangle
\cdot \vec{\psi}_j = \lambda_i \delta_{ij},
$$
where the dyadic $\langle \vec{d} \, \vec{d}^* \rangle$
is $\vec{\vec{C}}$, the correlation matrix of the data vector: 
$(\vec{d} \, \vec{d}^*)_{ij}\equiv d_i d^*_j$.
Now, the effect of operating this matrix on one of the
$\vec{\psi}_i$ must be expandable in terms of the complete
set, which shows that the $\smash{\vec{\psi}_j}$ must be the eigenvectors
of the correlation matrix:
$$
\langle \vec{d} \, \vec{d}^* \rangle \cdot
\vec{\psi}_j = \lambda_j \vec{\psi}_j.
$$

Vogeley \& Szalay further show that these uncorrelated modes are
optimal for representing the data: if the modes are
arranged in order of decreasing $\lambda$, and the
series expansion truncated after $n$ terms, the rms
truncation error is minimized for this choice of
eigenmodes. To prove this, consider
the truncation error
$$
\vec\epsilon = \vec{d} - \sum_{i=1}^n a_i \vec{\psi}_i = 
\sum_{i=n+1}^\infty a_i \vec{\psi}_i.
$$
The square of this is
$$
\langle \epsilon^2 \rangle = \sum_{i=n+1}^\infty \langle 
|a_i|^2 \rangle,
$$
where $\langle 
|a_i|^2 \rangle = \vec{\psi}_i^* \cdot \vec{\vec{C}} \cdot \vec{\psi}_i$,
as before. We want to minimize $\langle \epsilon^2 \rangle$
by varying the $\vec{\psi}_i$, but we need to do this in
a way that preserves normalization. This is achieved
by introducing a Lagrange multiplier, and minimizing
$$
\sum \vec{\psi}_i^* \cdot \vec{\vec{C}} \cdot \vec{\psi}_i
+ \lambda (1-\vec{\psi}_i^* \cdot \vec{\psi}_i).
$$
This is easily solved if we consider the more general
problem where $\vec{\psi}_i^*$ and $\vec{\psi}_i$ are
independent vectors:
$$
\vec{\vec{C}} \cdot \vec{\psi}_i = \lambda \psi_i.
$$
In short, the eigenvectors of $\vec{\vec{C}}$
are optimal in a least-squares sense for expanding the data.
The process of truncating the expansion is a form of
lossy \key{data compression}, since the size of the data vector
can be greatly reduced without significantly
affecting the fidelity of the resulting representation of the universe.

The process of diagonalizing the covariance
matrix of a set of data also goes by the more
familiar name of \key{principal components analysis},
so what is the difference between the KL
approach and PCA? In the above discussion, they are identical,
but the idea of choosing an optimal
eigenbasis is more general than PCA.
Consider the case where the covariance matrix
can be decomposed into a `signal' and a 
`noise' term:
$$
\vec{\vec{C}} =
\vec{\vec{S}} +
\vec{\vec{N}},
$$
where $\vec{\vec{S}}$ depends on cosmological
parameters that we might wish to estimate,
whereas $\vec{\vec{N}}$ is some fixed property of
the experiment under consideration.
In the simplest imaginable case, $\vec{\vec{N}}$
might be a diagonal matrix, so PCA diagonalizes
both $\vec{\vec{S}}$ and $\vec{\vec{N}}$.
In this case, ranking the PCA modes by eigenvalue would 
correspond to ordering the modes according to
signal-to-noise ratio. Data compression by truncating
the mode expansion then does the sensible thing: it rejects
all modes of low signal-to-noise ratio.

However, in general these matrices will not commute, and 
there will not be a single set of eigenfunctions that
are common to the $\vec{\vec{S}}$ and $\vec{\vec{N}}$
matrices. Normally, this would be taken to mean that it
is impossible to find a set of coordinates in which 
both are diagonal. This conclusion can however be evaded, as follows.
When considering the effect of coordinate transformations on
vectors and matrices, we are normally forced to consider
only rotation-like transformations that preserve the norm of
a vector (e.g. in quantum mechanics, so that states stay normalized).
Thus, we write $\vec{d}' = \vec{\vec{R}}\cdot \vec{d}$, where
$\vec{\vec{R}}$ is unitary, so that $\vec{\vec{R}} \cdot \vec{\vec{R}}^\dagger =
\vec{\vec{I}}$. If $\vec{\vec{R}}$ is chosen so that its columns are
the eigenvalues of $\vec{\vec{N}}$, then the transformed noise
matrix, $\smash{\vec{\vec{R}} \cdot \vec{\vec{N}} \cdot \vec{\vec{R}}^\dagger}$, 
is diagonal. Nevertheless, if the transformed $\vec{\vec{S}}$
is not diagonal, the two will not commute. This apparently insuperable
problem can be solved by using the fact that the data vectors are
entirely abstract at this stage. There is therefore no reason not to
consider the further transformation of scaling the data, so that
$\vec{\vec{N}}$ becomes proportional to the identity matrix. This
means that the transformation is no longer unitary -- but there is
no physical reason to object to a change in the normalization
of the data vectors.

Suppose we therefore make a further transformation
$$
\vec{d}'' = \vec{\vec{W}}\cdot \vec{d}'
$$
The matrix $\vec{\vec{W}}$ is related to the rotated noise matrix:
$$
\vec{\vec{N}}' = {\rm diag}\, (n_1,n_2,\dots) \quad \Rightarrow \quad
\vec{\vec{W}} = {\rm diag}\, (1/\sqrt{n_1},1/\sqrt{n_2},\dots).
$$
This transformation is termed \key{prewhitening} by Vogeley \& Szalay (1996),
since it converts the noise matrix to white noise,
in which each pixel has a unit noise that is uncorrelated with
other pixels. 
The effect of this transformation on the full covariance matrix is
$$
C_{ij}'' \equiv \langle d_i'' d_j''{}^* \rangle \quad \Rightarrow \quad
\vec{\vec{C}}'' = (\vec{\vec{W}}\cdot \vec{\vec{R}}) \cdot \vec{\vec{C}} \cdot
(\vec{\vec{W}}\cdot \vec{\vec{R}})^\dagger
$$
After this transformation, the noise and signal
matrices certainly do commute, and the optimal modes for
expanding the new data are once again the PCA
eigenmodes in the new coordinates:
$$
\vec{\vec{C}}''\cdot \vec{\psi}_i'' = \lambda \vec{\psi}_i''.
$$
These eigenmodes must be expressible
in terms of some modes in the original coordinates, $\vec{e}_i$:
$$
\vec{\psi}_i'' = (\vec{\vec{W}}\cdot \vec{\vec{R}}) \cdot \vec{e}_i.
$$
In these terms, the eigenproblem is 
$$
(\vec{\vec{W}}\cdot \vec{\vec{R}}) \cdot \vec{\vec{C}} \cdot
(\vec{\vec{W}}\cdot \vec{\vec{R}})^\dagger \cdot (\vec{\vec{W}}\cdot \vec{\vec{R}}) \cdot \vec{e}_i
= \lambda (\vec{\vec{W}}\cdot \vec{\vec{R}}) \cdot \vec{e}_i.
$$
This can be simplified using $\vec{\vec{W}}^\dagger \cdot \vec{\vec{W}} = \vec{\vec{N}}'{}^{-1}$
and $\vec{\vec{N}}'{}^{-1} = \vec{\vec{R}} \cdot \vec{\vec{N}}^{-1} \vec{\vec{R}}^\dagger$,
to give
$$
\vec{\vec{C}} \cdot \vec{\vec{N}}^{-1} \cdot \vec{e}_i = \lambda \vec{e}_i,
$$
so the required modes are eigenmodes of  $\vec{\vec{C}} \cdot \vec{\vec{N}}^{-1}$.
However, care is required when considering the orthonormality of the $\vec{e}_i$:
$\smash{\vec{\psi}_i^\dagger \cdot \vec{\psi}_j = \vec{e}_i^\dagger \cdot \vec{\vec{N}}^{-1} \cdot \vec{e}_j}$,
so the $\vec{e}_i$ are not orthonormal. If we write $\vec{d} = \sum_i a_i \vec{e}_i$, then
$$
a_i = (\vec{\vec{N}}^{-1} \cdot \vec{e}_i)^\dagger \cdot \vec{d} \equiv \vec{\psi}_i^\dagger \cdot \vec{d}.
$$ 
Thus, the modes used to
extract the compressed data by dot product satisfy
$\vec{\vec{C}} \cdot \vec{\psi} = \lambda \vec{\vec{N}}\cdot \vec{\psi}$, or finally
$$
\vec{\vec{S}}
\cdot \vec{\psi} = \lambda\, \vec{\vec{N}} \cdot \vec{\psi},
$$
given a redefinition of $\lambda$.
The optimal modes are thus eigenmodes of
$\vec{\vec{N}}^{-1} \cdot \vec{\vec{S}}$,
hence the name \key{signal-to-noise eigenmodes}
(Bond 1995; Bunn 1996).

It is interesting to appreciate that the set of KL modes
just discussed is also the `best'
set of modes to choose from a completely different
point of view: they are the modes that
are optimal for estimation of a parameter via
maximum likelihood.
Suppose we write the compressed data vector, $\vec{x}$, in terms of a
non-square matrix $\vec{\vec{A}}$ (whose rows are the basis
vectors $\vec{\psi}_i^*$):
$$
\vec{x}=\vec{\vec{A}}\cdot \vec{d}.
$$
The transformed covariance matrix is
$$
\vec{\vec{D}} \equiv \langle \vec{x}\vec{x}^\dagger \rangle
=
\vec{\vec{A}} \cdot \vec{\vec{C}} \cdot \vec{\vec{A}}^\dagger.
$$
For the case where the original data obeyed Gaussian
statistics, this is true for the compressed data also, so the
likelihood is
$$
-2\ln {\cal L} = \ln {\rm det}\, \vec{\vec{D}} + 
\vec{x}^* \cdot \vec{\vec{D}}^{-1} \cdot \vec{x} + {\rm constant}
$$
The normal variance on some parameter $p$ (on
which the covariance matrix depends) is
$$
{1\over \sigma_p^2} = {d^2 [-2\ln {\cal L}] \over dq^2}.
$$
Without data, we don't know this, so it is common to use
the expectation value of the rhs as
an estimate (recently, there has been a tendency to dub this
the `Fisher matrix').

We desire to optimize $\sigma_p$ by an appropriate choice
of data-compression vectors, $\vec{\psi}_i$.
By writing $\sigma_p$ in terms of $\vec{\vec{A}}$,
$\vec{\vec{C}}$ and $\vec{d}$, it may eventually be
shown that the desired optimal modes satisfy
$$
\left({d\over dp}\, \vec{\vec{C}} \right)
\cdot \vec{\psi} = \lambda\, \vec{\vec{C}} \cdot \vec{\psi}.
$$
For the case where the parameter of interest is
the cosmological power, the matrix on the lhs is 
just proportional to $\vec{\vec{S}}$, so we have to solve
the eigenproblem
$$
\vec{\vec{S}}
\cdot \vec{\psi} = \lambda\, \vec{\vec{C}} \cdot \vec{\psi}.
$$
With a redefinition of $\lambda$, this becomes
$$
\vec{\vec{S}}
\cdot \vec{\psi} = \lambda\, \vec{\vec{N}} \cdot \vec{\psi}.
$$
The optimal modes for parameter estimation in the
linear case are thus identical to the PCA modes of the
prewhitened data discussed above.
The more general expression was given by
Tegmark, Taylor \& Heavens (1997), and it is
only in this case, where the covariance matrix
is not necessarily linear in the parameter of interest, that
the KL method actually differs from PCA.

The reason for going to all this trouble is that the likelihood
can now be evaluated much more rapidly, using the compressed data.
This allows extensive model searches over large parameter spaces
that would be unfeasible with the original data (since inversion
of an $N\times N$ covariance matrix takes a time proportional
to $N^3$). Note however that the price paid for this
efficiency is that a different set of modes need to be
chosen depending on the model of interest, and that these
modes will not in general be optimal for expanding the
dataset itself.
Nevertheless, it may be expected that application of these methods will
inevitably grow as datasets increase in size. Present applications
mainly prove that the techniques work: see 
Matsubara, Szalay \& Landy (1999) for application to the LCRS, or
Padmanabhan, Tegmark \&  Hamilton (1999) for the UZC survey.
The next generation of experiments will probably be
forced to resort to data compression of this sort,
rather than using it as an elegant alternative method
of analysis.

\sec{Redshift-space effects}

Peculiar velocity fields are responsible for the
distortion of the clustering pattern in 
redshift space, as first clearly articulated
by Kaiser (1987). 
For a survey that subtends a small angle
(i.e. in the \key{distant-observer approximation}), a good approximation to
the anisotropic redshift-space Fourier spectrum is given
by the Kaiser function together with a damping
term from nonlinear effects:
$$
\delta_k^s=\delta_k^r (1+\beta\mu^2)D(k\sigma\mu),
$$
where $\beta=\Omega_m^{0.6}/b$, $b$ being the
linear bias parameter of the galaxies under study,
and $\mu={\bf \hat k \cdot \hat r}$.
For an exponential distribution of relative
small-scale peculiar velocities (as seen empirically),
the damping function is 
$D(y)\simeq (1+y^2/2)^{-1/2}$, and $\sigma\simeq 400\kms$
is a reasonable estimate for the pairwise velocity dispersion of galaxies
(e.g. Ballinger, Peacock \& Heavens 1996).

In principle, this distortion should be a robust
way to determine $\Omega$ (or at least $\beta$).
In practice, the effect has not been easy to see
with past datasets. This is mainly a question of
depth: a large survey is needed in order to beat down
the shot noise, but this tends to favour bright
spectroscopic limits. This limits the result both because
relatively few modes in the linear regime are
sampled, and also because local survey volumes
will tend to violate the small-angle approximation.
Strauss \& Willick (1995) and Hamilton (1997a) review
the practical application of redshift-space distortions.
In the next section, preliminary results are presented
from the 2dF redshift survey, which shows the
distortion effect clearly for the first time.

\sec{The state of the art in LSS}

\ssec{The APM survey}

In the past few years, much attention has been attracted by
the estimate of the galaxy power spectrum from the APM
survey (Baugh \& Efstathiou 1993, 1994; Maddox et al. 1996).
The APM result was generated from
a catalogue of $\sim 10^6$ galaxies derived from UK
Schmidt Telescope photographic plates scanned with the
Cambridge Automatic Plate Measuring machine; because it is based
on a \key{deprojection} of angular clustering, it is immune to the
complicating effects of redshift-space distortions.
The difficulty, of course, is in ensuring that any
low-level systematics from e.g. spatial variations
in magnitude zero point are sufficiently well
controlled that they do not mask the cosmological
signal, which is of order $w(\theta) \ls 0.01$ 
at separations of a few degrees.

The best evidence that the
APM survey has the desired uniformity is the \key{scaling test}, 
where the correlations in fainter magnitude slices are expected to 
move to smaller scales and be reduced in amplitude.
If we increase the depth of the survey by some
factor $D$, the new angular correlation function will be
$$
w'(\theta) = {1\over D} \, w(D\theta).
$$
The APM survey passes this test well; once the overall 
redshift distribution is known, it is possible to obtain
the spatial power spectrum by inverting a convolution integral:
$$
w(\theta)= \int_0^\infty y^4\phi^2\, dy\ \int_0^\infty
  \pi\, \Delta^2(k)\, J_0(ky\theta)\, dk/k^2
$$
(where zero spatial curvature is assumed).
Here, $\phi(y)$ is the comoving density at comoving
distance $y$, normalized so that $\int y^2\phi(y)\, dy=1$.

This integral was inverted numerically by Baugh \& Efstathiou
(1993), and gives an impressively accurate determination
of the power spectrum.
The error estimates are derived empirically from the scatter
between independent regions of the sky, and so should be
realistic. If there are no undetected systematics, these
error bars say that the power is very accurately determined.
The APM result has been investigated
in detail by a number of authors (e.g. 
Gazta\~naga \& Baugh 1998; Eisenstein \& Zaldarriaga 1999)
and found to be robust;
this has significant implications if true.

\ssec{Past redshift surveys}

Because of the sheer number of galaxies, plus the large
volume surveyed, the APM survey outperforms redshift
surveys of the past, at least for the purpose of
determining the power spectrum. The largest surveys
of recent years (CfA: Huchra et al. 1990; 
LCRS: Shectman et al. 1996; PSCz: Saunders et al. 1999)
contain of order $10^4$ galaxy redshifts, and their
statistical errors are considerably larger than
those of the APM. On the other hand, it is of
great importance to compare the results of deprojection
with clustering measured directly in 3D.

This comparison was carried out
by Peacock \& Dodds (1994; PD94).
The exercise is not straightforward, because the 3D
results are affected by redshift-space distortions;
also, different galaxy tracers can be biased to 
different extents. The approach taken was to use each
dataset to reconstruct an estimate of 
the linear spectrum, allowing the relative bias factors
to float in order to make these estimates agree
as well as possible
(figure 5). To within a scatter of perhaps a factor 1.5 in
power, the results were consistent with a $\Gamma\simeq 0.25$ CDM model.
Even though the subsequent sections will discuss some possible
disagreements with the CDM models at a higher level of
precision, the general existence of CDM-like curvature
in the spectrum is likely to be an important clue to the
nature of the dark matter.

\japfig{37}{117}{490}{749}{japfig5.eps}
{The PD94 compilation of power-spectrum
measurements. The upper panel shows raw
power measurements; the lower shows these data
corrected for relative bias, nonlinear effects, and
redshift-space effects.}

\ssec{The 2dF survey}

The proper resolution of many of the observational questions
regarding the large-scale distribution of galaxies requires
new generations of redshift survey that push beyond
the $N=10^5$ barrier. Two groups are pursuing this
goal. The Sloan survey (e.g. Margon 1999) is using a
dedicated 2.5-m telescope to measure redshifts for 
approximately 700,000 galaxies to $r=18.2$ in the North Galactic Cap.
The 2dF survey (e.g. Colless 1999) is using a fraction of the
time on the 3.9-m Anglo-Australian Telescope plus
Two-Degree Field spectrograph to measure 250,000 galaxies
from the APM survey to $B_J=19.45$ in the South Galactic Cap.
At the time of writing, the Sloan spectroscopic
survey has yet to commence. However, the 2dF project has measured 77,000
redshifts, and some preliminary clustering results 
are given below.
For more details of the survey, particularly the team
members whose hard work has made all this possible,
see
{\tt http://www.mso.anu.edu.au/2dFGRS/}.

\japfig{57}{211}{515}{588}{japfig6.eps}
{A 4-degree thick slice of the Southern
strip of the 2dF redshift survey. This restricted region 
alone contains 16,419 galaxies.}

One of the advantages of 2dF is that it is a fully sampled survey,
so that the space density out to the depth imposed by the
magnitude limit (median $z=0.12$) is as high as nature
allows: apart from a tail of low surface brightness 
galaxies (inevitably omitted from any spectroscopic survey),
the 2dF measure all the galaxies that exist over a
cosmologically representative volume. It is the first to
achieve this goal. The fidelity of the resulting map of
the galaxy distribution can be seen in figure 6, which shows
a small subset of the data: a slice of thickness 4
degrees, centred at declination $-27^\circ$.

An issue with using the 2dF data in their current
form is that the sky has to be divided into circular `tiles'
each two degrees in diameter (`2dF' = `two-degree field', within
which the AAT is able to measure 400 spectra simultaneously;
see {\tt http://www.aao.gov.au/2df/} for details of the instrument).
The tiles are positioned adaptively, so that larger overlaps
occur in regions of high galaxy density. It this way, it is
possible to place a fibre on $>95\%$ of all galaxies.
However, while the survey is in progress, there exist parts
of the sky where the overlapping tiles have not yet been observed,
and so the effective sampling fraction is only $\simeq 50\%$.
These effects can be allowed for in two different ways.
In clustering analyses, we compare the counts of pairs
(or $n$-tuplets) of galaxies in the data to the corresponding
counts involving an unclustered random catalogue. The
effects of variable sampling can therefore be dealt with
either by making the density of random points fluctuate
according to the sampling, or by weighting observed galaxies
by the reciprocal of the sampling factor for the zone in
which they lie. The former approach is better from the
point of view of shot noise, but the latter may be safer
if there is any suspicion that the sampling fluctuations
are correlated with real structure on the sky. In practice,
both strategies give identical answers for the results below.

At the two-point level, the most direct quantity to compute
is the \key{redshift-space correlation function}. This is an
anisotropic function of the orientation of a galaxy 
pair, owing to peculiar velocities. We therefore evaluate
$\xi$ as a function of 2D  separation in terms of coordinates
both parallel and perpendicular to the line of sight.
If the comoving radii of two galaxies
are $y_1$ and $y_2$ and their total separation is $r$, then
we define coordinates
$$
\pi \equiv |y_1-y_2|; \quad\quad \sigma = \sqrt{r^2-\pi^2}.
$$
The correlation function measured in these coordinates
is shown in figure 7.
In evaluating $\xi(\sigma, \pi)$, the optimal
radial weight discussed above has been applied, so that
the noise at large $r$ should be representative of
true cosmic scatter.

\begin{figure*}
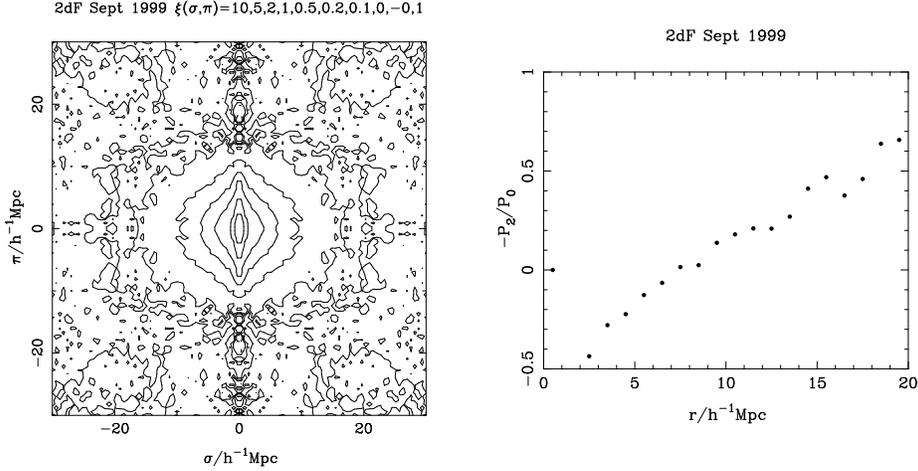

\centering\mbox{\epsfxsize=0.45\hsize
\epsfbox[38 123 521 658]{japfig7a.eps} \quad\quad
\raise 1.5em\hbox{
\epsfxsize=0.45\hsize
\epsfbox[60 211 511 633]{japfig7b.eps}
}
}
\caption[]{The redshift-space correlation function from the
2dF data, $\xi(\sigma, \pi)$, with a bin size of
$0.6\mpcoh$.
$\sigma$ is the pair separation transverse to the line
of sight; $\pi$ is the radial separation.
This plot clearly displays redshift distortions, with
`fingers of God' at small scales and the coherent
Kaiser squashing at large $\sigma$.
The distortions are quantified via the 
quadrupole-to-monopole ratio of $\xi$ as a function
of radius in the second panel. The contours are
round at $r=7\mpcoh$, but flatten progressively thereafter.
}
\end{figure*}

The correlation-function results display very clearly
the two signatures of redshift-space distortions discussed
above. The \key{fingers of God} from small-scale random
velocities are very clear, as indeed has been the case
from the first redshift surveys (e.g. Davis \& Peebles 1983).
However, this is arguably the first time that the large-scale
flattening from coherent infall has been really obvious in the
data.

A good way to quantify the flattening is to analyze the
clustering as a function of angle into Legendre polynomials:
$$
\xi_\ell(r) = {2\ell+1\over 2} \int_{-1}^1 \xi(\sigma=r\sin\theta,
\pi=r\cos\theta)\; P_\ell(\cos\theta)\; d\cos\theta.
$$
The quadrupole-to-monopole ratio should be a clear indicator
of coherent infall. In linear theory, it is given by
$$ 
{\xi_2\over \xi_0} = f(n) \,
{4\beta/3 + 4\beta^2/7 \over
1 + 2\beta/3 + \beta^2/5 },
$$
where $f(n)=(3+n)/n$  (Hamilton 1992). On small and intermediate
scales, the effective spectral index is negative,
so the quadrupole-to-monopole
ratio should be negative, as observed.

However, it is clear that the results on the
largest scales are still significantly affected by
finger-of-God smearing. The best way to interpret the
observed effects is to calculate the same quantities
for a model. To achieve this, we use the observed
APM 3D power spectrum, plus the distortion model
discussed above. This gives the plots shown in figure 8.
The free parameter is $\beta$, and this is set at
a value of 0.5, approximately consistent with
other arguments for a universe with $\Omega=0.3$
and little large-scale bias (e.g. Peacock 1997).
Although a quantitative comparison has not yet
been carried out, it is clear that this plot
closely resembles the observed data.

\begin{figure*}
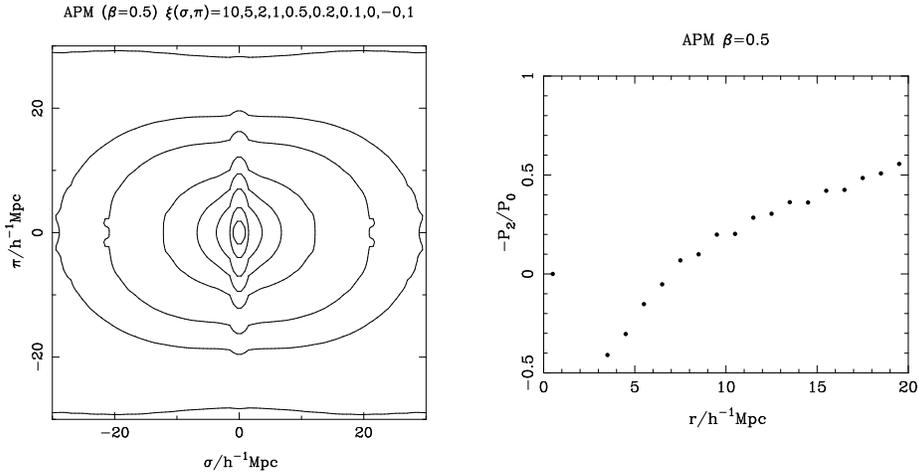

\centering\mbox{\epsfxsize=0.45\hsize
\epsfbox[38 123 521 658]{japfig8a.eps} \quad\quad
\raise 1.5em\hbox{
\epsfxsize=0.45\hsize
\epsfbox[60 211 511 633]{japfig8b.eps}
}
}
\caption[]{The redshift-space correlation function predicted from the
real-space APM power spectrum, assuming the model of
Ballinger, Peacock \& Heavens (1996), with $\beta=0.5$.}
\end{figure*}

By the end of 2001, the size of the 2dF survey should have expanded
by a factor 3, increasing the pair counts tenfold. It should
then be possible to trace the correlations well beyond the present
limit, and follow the redshift-space distortion well into the
linear regime.
However, the biggest advantage of a survey of this size and
uniformity is the ability to subdivide it. All analyses to date
have lumped together very different kinds of galaxies, whereas
we know from morphological segregation that different classes
of galaxy have spatial distributions that differ from each other.
The homogeneous 2dF data allow classification into
different galaxy types (representing, physically, a sequence
of star-formation rates), from the spectra alone
(Folkes et al. 1999). It will be a critical test to see if
the distortion signature can be picked up in each type
individually. Although the large-scale behaviour of each
galaxy type will probably be quite similar, differences in
the clustering properties will inevitably arise on smaller
scales, giving important information about
the sequence of galaxy formation.

\sec{Small-scale clustering}

\ssec{History}

One of the earliest models to be used to interpret the
galaxy correlation function was to consider a density
field composed of randomly-placed independent clumps
with some universal density profile (Neyman, Scott \& Shane 1953; Peebles 1974).
Since the clumps are placed at random, the only correlations
arise from points in the same clump.
The correlations are easily deduced by using statistical
isotropy: calculate the excess number of pairs separated
by a distance $r$ in the $z$ direction (chosen as some arbitrary
polar axis in a spherically-symmetric clump).
For power-law clumps, with $\rho= n B r^{-\epsilon}$, truncated at
$r=R$, this model gives $\xi\propto r^{3-2\epsilon}$
in the limit $r\ll R$, provided $3/2 < \epsilon <3$.
Values $\epsilon >3$ are unphysical, and
require a small-scale cutoff to the profile. There is no such objection
to $\epsilon < 3/2$, and the expression for $\xi$ tends to a
constant for small $r$ in this case (see Yano \& Gouda 1999).

A long-standing problem for this model is that the
correlation function in this case is much flatter than is
observed for galaxies: $\xi \propto r^{-1.8}$ is the
canonical slope, requiring $\epsilon=2.4$.
The first reaction may be to say that the
model is incredibly naive by comparison with our
sophisticated present understanding of the nonlinear
evolution of CDM density fields. However, as will
be shown below, it may after all contain more than a grain of
truth.

\ssec{The CDM clustering problems}

A number of authors have pointed out that the detailed
spectral shape inferred from galaxy data 
appears to be inconsistent with that of nonlinear
evolution from CDM initial conditions.
(e.g. Efstathiou,  Sutherland \& Maddox 1990;
Klypin, Primack \& Holtzman 1996; Peacock 1997).
Perhaps the most detailed work was carried out by the
VIRGO consortium, who carried out $N=256^3$ simulations of
a number of CDM models (Jenkins et al. 1998). Their results
are shown in figure 9, which gives the nonlinear power
spectrum at various times (cluster normalization is chosen
for $z=0$) and contrasts this with the APM data.
The lower small panels are the scale-dependent bias
that would  required if the model did in fact describe the real universe,
defined as
$$
b(k)\equiv \left({\Delta^2_{\rm gals}(k)\over\Delta^2_{\rm mass}}\right)^{1/2}.
$$
In all cases, the required bias is non-monotonic; it rises at
$k\gs 5\mpcoh$, but also displays a bump around
$k\simeq 0.1\mpcoh$.
If real, this feature seems impossible to understand as a
genuine feature of the mass power spectrum; certainly, it is not at
a scale where the effects of even a large baryon fraction
would be expected to act (Eisenstein et al. 1998; Meiksin, White
\& Peacock 1999).

\japfig{20}{75}{525}{700}{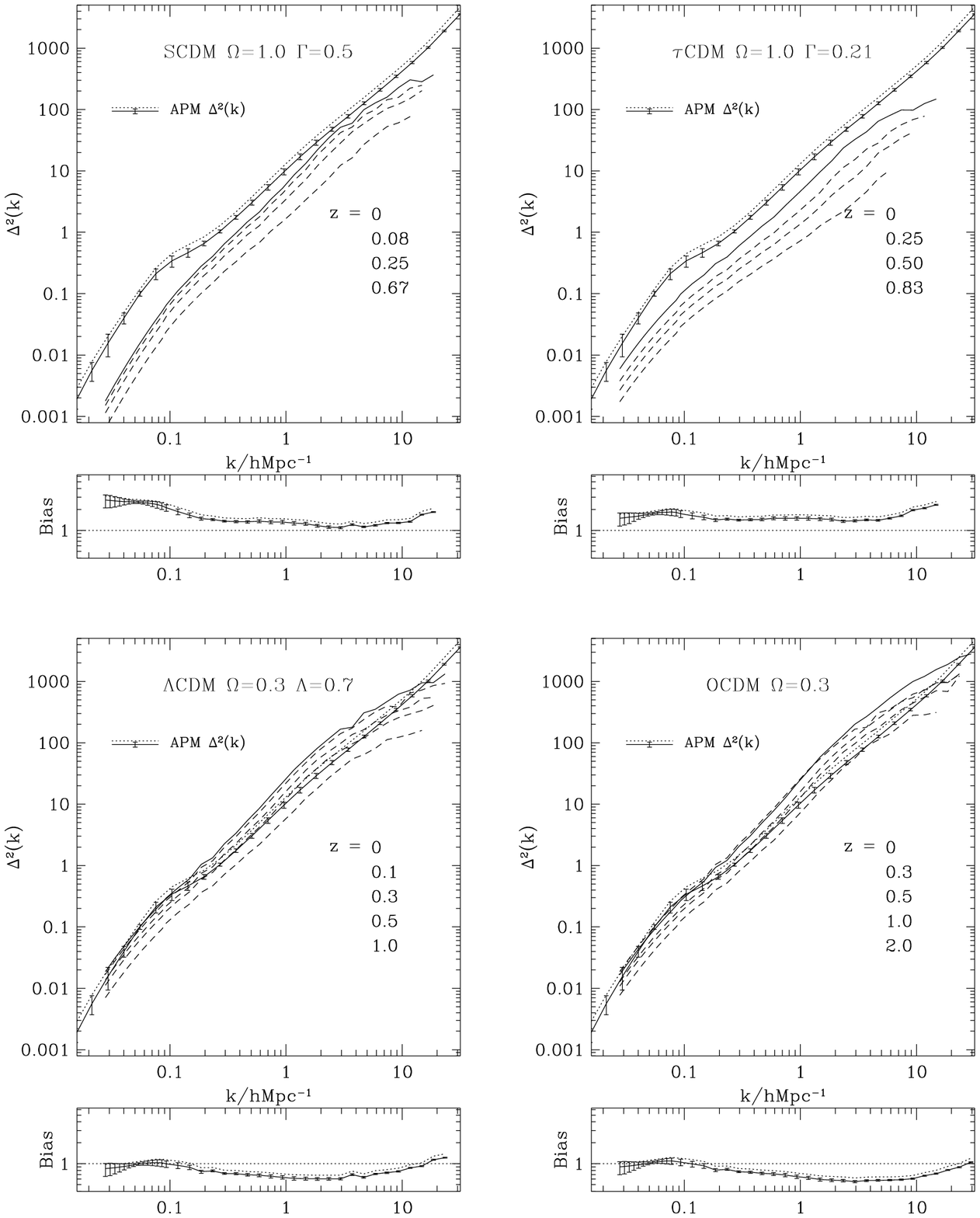}
{The nonlinear evolution of various CDM
power spectra, as determined by the Virgo
consortium (Jenkins et al. 1998).}

\sec{Bias}

The conclusions from the above discussion are either that
the physics of dark matter and structure formation are
more complex than in CDM models, or that the relation
between galaxies and the overall matter distribution
is sufficiently  complicated that the effective bias is
not a simple slowly-varying monotonic function of position.

\ssec{Simple bias models}

The simplest assumption is that all the complicated physical effects 
leading to galaxy formation depend in a causal (but nonlinear) way on the
local mass density, so that we write
$$
\rho_{\rm light}=f(\rho_{\rm mass}).
$$
Coles (1993) showed that, under rather general assumptions, this
equation would lead to an effective bias that was a monotonic
function of scale. This issue was investigated in some detail
by Mann, Peacock \& Heavens (1998), who verified Coles'
conclusion in practice for simple few-parameter forms for
$f$, and found in all cases that the effective bias varied
rather weakly with scale. The APM results thus are either inconsistent
with a CDM universe, or require non-local bias.

\japfig{0}{0}{484}{472}{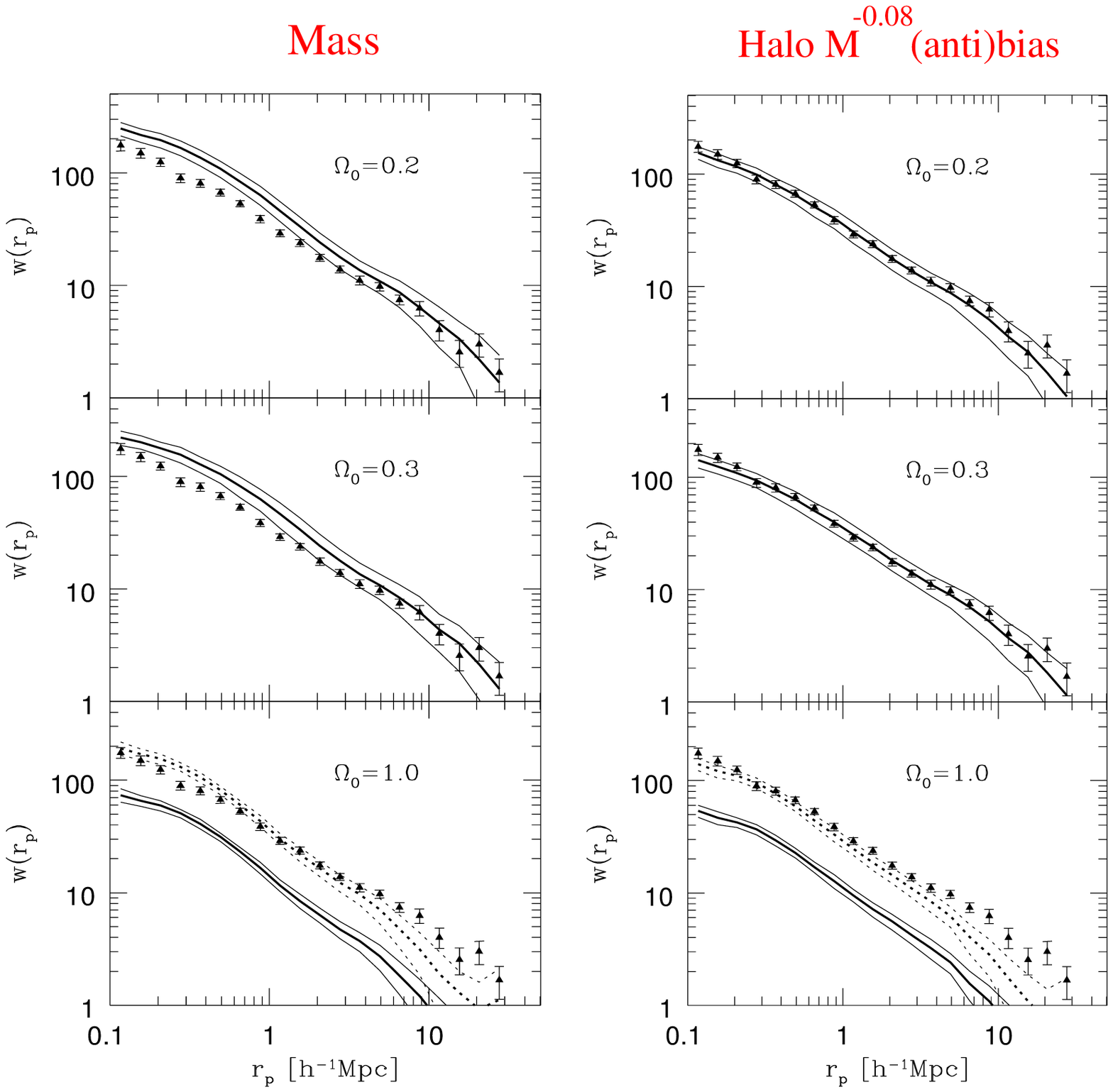}
{The projected correlation function from the
LCRS fails to match CDM models when comparison is made to
just the mass distribution. However, the agreement
is excellent when allowance is made for a small
degree of scale-dependent antibias; galaxy formation
is suppressed in the most massive haloes
(Jing, Mo \& B\"orner 1998).}

A puzzle with regard to this conclusion is provided
by the work of Jing, Mo \& B\"orner (1998). They
evaluated the projected real-space correlations
for the LCRS survey (see figure 10). This statistic also
fails to match the prediction of CDM models, but this
can be amended by introducing a simple {\it antibias\/}
scheme, in which galaxy formation is suppressed in the
most massive haloes. This scheme should in practice
be very similar to the Mann, Peacock \& Heavens recipe
of a simple weighting of particles as a function of the
local density; indeed, the main effect is a change of
amplitude, rather than shape of the correlations.
The puzzle is this: if the APM power spectrum is used
to predict the projected correlation function, the
result agrees almost exactly with the LCRS. Either
projected correlations are a rather
insensitive statistic, or perhaps the Baugh \& Efstathiou
deconvolution procedure used to get $P(k)$ has
exaggerated the significance of features in the spectrum.
The LCRS results are one reason for treating the apparent conflict 
between APM and CDM with caution.

\ssec{Halo correlations}

In reality, bias is unlikely to be completely causal,
and this has led some workers to explore stochastic bias
models, in which
$$
\rho_{\rm light}=f(\rho_{\rm mass}) + \epsilon,
$$
where $\epsilon$ is a random field that is uncorrelated with the
mass density (Pen 1998; Dekel \& Lahav 1999).
Although truly stochastic effects are possible in galaxy formation,
a relation of the above form is expected when the
galaxy and mass densities are filtered on some scale
(as they always are, in practice). Just averaging a 
galaxy density that is a nonlinear
function of the mass will lead to some scatter when comparing with the
averaged mass field; a scatter will also arise when the
relation between mass and light is non-local, however, and this
may be the dominant effect.

The simplest and most important example of non-locality
in the galaxy-formation process is to recognize that
galaxies will generally form where there are galaxy-scale
haloes of dark matter. In the past, it was
generally believed that dissipative processes were
critically involved in galaxy formation, since pure
collisionless evolution would lead to the destruction
of galaxy-scale haloes when they are absorbed into the
creation of a larger-scale nonlinear system such as a group
or cluster. However, it turns out that this
{\it overmerging problem\/} was only an artefact of
inadequate resolution. When a simulation is carried out
with $\sim 10^6$ particles in a rich cluster, the cores of
galaxy-scale haloes can still be identified after many crossing
times (Ghigna et al. 1997). Furthermore, if catalogues of
these `sub-haloes' are created within a cosmological-sized
simulation, their correlation function is quite different from that
of the mass, resembling the single power law seen in galaxies 
(e.g. Klypin et al. 1999; Ma 1999).

These are very important results, and they
hold out the hope that many of the issues concerning
where galaxies form in the cosmic density field can be settled
within the domain of collisionless simulations. 
Dissipative physics will still be needed to understand in
detail the star-formation history within a galaxy-scale
halo. Nevertheless, the idea that there may be a one-to-one correspondence
between galaxies and galaxy-scale dark-matter haloes
is clearly an enormous simplification -- and one that increases
the chance of making robust predictions of the statistical
properties of the galaxy population.

\ssec{Numerical galaxy formation}

The formation of galaxies must be a non-local process to
some extent. The modern paradigm was introduced by White \& Rees (1978):
galaxies form through the cooling of baryonic material in
virialized haloes of dark matter. The virial radii of these
systems are in excess of 0.1~Mpc, so there is the potential
for large differences in the correlation properties of
galaxies and dark matter on these scales.

\japfig{18}{144}{574}{701}{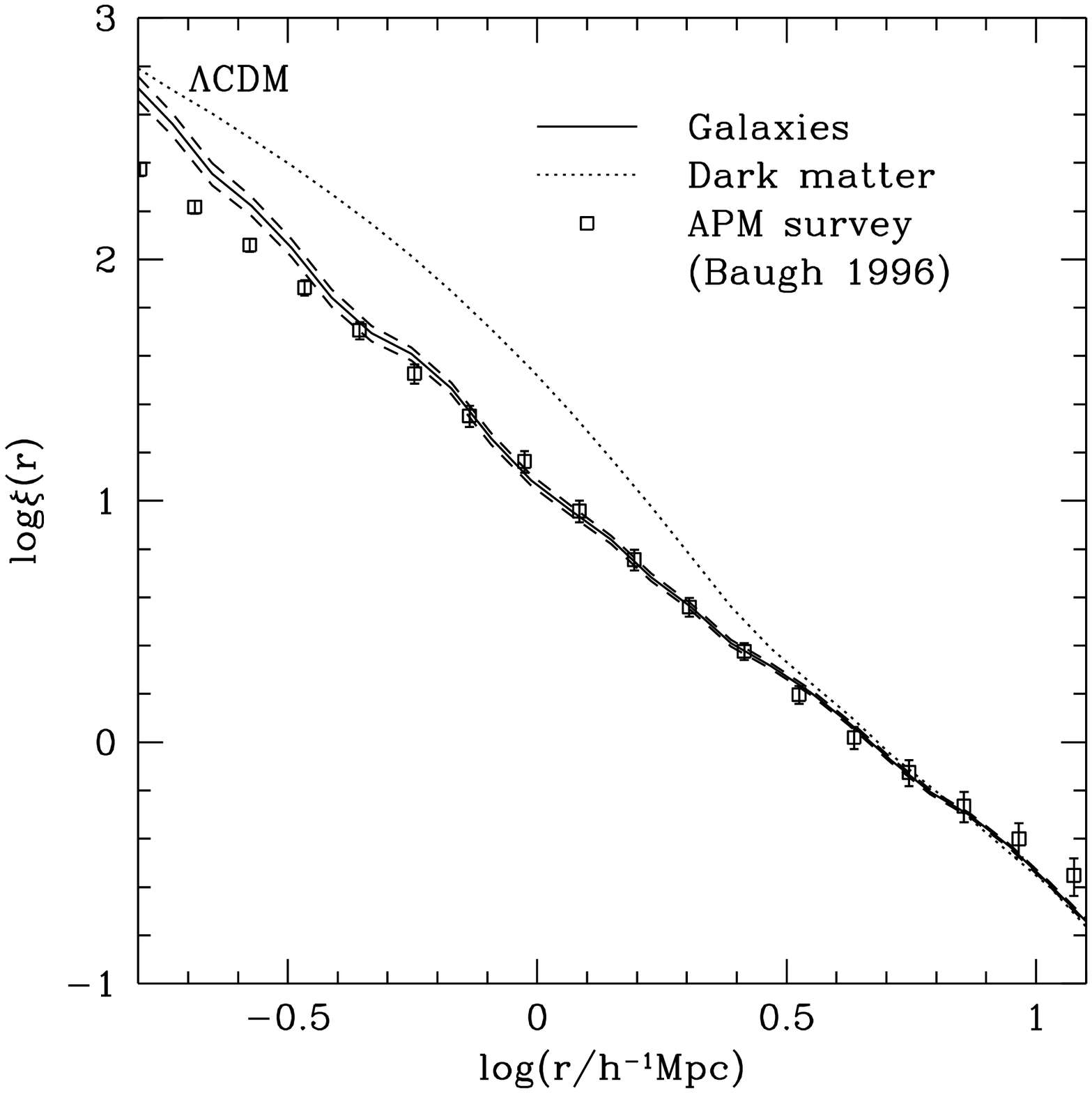}
{The correlation function of galaxies in the
semianalytical simulation of an LCDM
universe by Benson et al. (1999).}

A number of studies have indicated that the observed
galaxy correlations may indeed be reproduced by CDM models.
The most direct approach is a numerical simulation that
includes gas, and relevant dissipative processes.
This is challenging, but just starting to be feasible with
current computing power (Pearce et al. 1999). The alternative
is `semianalytic' modelling, in which the merging history
of dark-matter haloes is treated via the extended Press-Schechter
theory (Bond et al. 1991), and the location of galaxies within
haloes is estimated using dynamical-friction arguments
(e.g. Cole et al. 1996; Kauffmann et al. 1996; 
Somerville \& Primack 1997). Both these approaches
have yielded similar conclusions, and shown how CDM models
can match the galaxy data: specifically, the low-density
flat $\Lambda$CDM model that is favoured on other
grounds can yield a correlation function that is close to a
single power law over $1000 \gs \xi \gs 1$, even though the
mass correlations show a marked curvature over this range
(Pearce et al. 1999; Benson et al. 1999; see figure 11).
These results are impressive, yet it is frustrating to have a result
of such fundamental importance emerge from a complicated
calculational apparatus. 
There is thus some motivation for constructing a simpler
heuristic model that captures the main processes at work in
the full semianalytic models. The following section
describes an approach of this sort (Peacock \& Smith,
in preparation).

\ssec{Halo-ology and bias}

We mentioned above the early model of Neyman, Scott \& Shane (1953),
in which the nonlinear density field was taken to be a superposition
of randomly-placed clumps. With our present knowledge
about the evolution of CDM universes, we can
make this idealised model considerably more realistic:
hierarchical models are expected to
contain a distribution of masses of clumps, which
have density profiles that are more complicated than
isothermal spheres.  These issues are well
studied in $N$-body simulations, and highly accurate
fitting formulae exist, both for the mass function and
for the density profiles. 
Briefly, we use the mass function of
Sheth \& Tormen (1999; ST) and the halo profiles
of Moore et al. (1999; M99).
$$
\eqalign{
f(\nu) &= 0.21617[ 1 + (\sqrt{2}/\nu^2)^{0.3} ] \exp[-\nu^2/(2\sqrt{2})] \cr
\Rightarrow F(>\nu) &= 0.32218[1-\erf(\nu/2^{3/4})] \cr
& + 0.14765 \Gamma[0.2, \nu^2/(2\sqrt{2})], \cr
}
$$
where $\Gamma$ is the incomplete gamma function. 

Recently, it has been claimed by Moore et al. (1999; M99) that the
commonly-adopted density profile of Navarro, Frenk \& White
(1996; NFW) is in error at small $r$. M99 proposed the
alternative form
$$
\rho/\rho_b = {\Delta_c \over y^{3/2} (1+y^{3/2}) }; \quad (r<r_{\rm vir}); \quad y\equiv r/r_c.
$$
Using this model,
it is then possible to calculate the correlations
of the nonlinear density field, neglecting only the
large-scale correlations in halo positions. The
power spectrum determined in this way is shown in figure 12,
and turns out to agree very well with the
exact nonlinear result on small and intermediate scales.
The lesson here is that a good deal of the
nonlinear correlations of the dark matter field
can be understood as a distribution of random clumps,
provided these are given the correct distribution of
masses and mass-dependent density profiles.

\japfig{57}{211}{515}{588}{japfig12.eps}
{The power spectrum for the $\Lambda$CDM model.
The solid lines contrast the linear spectrum
with the nonlinear spectrum, calculated according to
the approximation of PD96. The spectrum according
to randomly-placed haloes is denoted by open circles;
if the linear power spectrum is added, the main
features of the nonlinear spectrum are well reproduced.}

How can we extend this model to understand how the
clustering of galaxies can differ from that of the mass?
There are two distinct ways in which a degree of bias is inevitable:

\japitem{(1)} Halo occupation numbers. For low-mass haloes, the
probability of obtaining an $L^*$ galaxy must fall to zero.
For haloes with mass above this lower limit, the number of
galaxies will in general not scale with halo mass.

\japitem{(2)} Nonlocality. Galaxies can orbit within their
host haloes, so the probability of forming a galaxy depends
on the overall halo properties, not just the density at a point.
Also, the galaxies will end up at special places within
the haloes: for a halo containing only one galaxy, the
galaxy will clearly mark the halo centre. In general,
we expect one central galaxy and a number of satellites.

The numbers of galaxies that form in a halo of a
given mass is the prime quantity that numerical models
of galaxy formation aim to calculate.
However, for a given assumed background cosmology, the
answer may be determined empirically.
Galaxy redshift surveys have been analyzed via grouping
algorithms similar to the `friends-of-friends' method
widely employed to find virialized clumps in $N$-body
simulations. With an appropriate correction for the
survey limiting magnitude, the observed number of galaxies in
a group can be converted to an estimate of the total
stellar luminosity in a group. This allows a
determination of the All Galaxy System (AGS)
luminosity function: the distribution of virialized
clumps of galaxies as a function of their total
luminosity, from small systems like the Local Group to
rich Abell clusters.

\japfig{57}{211}{515}{588}{japfig13.eps}
{The empirical luminosity--mass relation
required to reconcile the observed AGS luminosity function
with two variants of CDM. $L^*$ is the characteristic
luminosity in the AGS luminosity function
($L^* = 7.6\times 10^{10}h^{-2} L_\odot$).
Note the rather flat slope around
$M=10^{13}$ to $10^{14}h^{-1}M_\odot$,
especially for $\Lambda$CDM.}

The AGS function for the CfA survey was investigated by
Moore, Frenk \& White (1993), who found that the
result in blue light was well described by
$$
d\phi = \phi^*\, \left[ (L/L^*)^\beta + (L/L^*)^\gamma \right]^{-1}\;
dL/L^*,
$$
where $\phi^*=0.00126h^3\rm Mpc^{-3}$, $\beta=1.34$, $\gamma=2.89$;
the characteristic luminosity is $M^*=-21.42 + 5\log_{10}h$ in
Zwicky magnitudes, corresponding to $M_B^*=-21.71 + 5\log_{10}h$,
or $L^*  = 7.6\times 10^{10}h^{-2} L_\odot$, assuming
$M_B^\odot=5.48$.
One notable feature of this function is that it is
rather flat at low luminosities, in contrast to the
mass function of dark-matter haloes (see Sheth \& Tormen 1999).
It is therefore clear that any fictitious galaxy catalogue
generated by randomly sampling the mass is unlikely to be a
good match to observation.
The simplest cure for this deficiency is to assume that the
stellar luminosity per virialized halo is a monotonic, but nonlinear,
function of halo mass. The required luminosity--mass
relation is then easily deduced by finding the luminosity
at which the integrated AGS density $\Phi(>L)$ matches the
integrated number density of haloes with mass $>M$.
The result is shown in figure 13.

We can now return to the halo-based galaxy power spectrum
and use the correct occupation number, $N$, as a function of 
mass. This is needs a little care at small numbers,
however, since the number of haloes with occupation number unity
affects the correlation properties strongly. These
haloes contribute no correlated pairs, so they simply
dilute the signal from the haloes with $N\ge 2$. The existence
of antibias on intermediate scales can probably be traced to
the fact that a large fraction of galaxy groups contain only
one $>L_*$ galaxy. Finally, we need to put the
galaxies in the correct location, as discussed above.
If one galaxy always occupies the halo centre, with others
acting as satellites, the small-scale correlations automatically
follow the slope of the halo density profile, which keeps them
steep. The results of this exercise are shown in figure 14.

\japfig{57}{211}{515}{588}{japfig14.eps}
{The power spectrum for a galaxy catalogue constructed from
the $\Lambda$CDM model. A reasonable
agreement with the APM data (solid line) is achieved by
simple empirical adjustment of the occupation number
of galaxies as a function of halo mass, plus a
scheme for placing the haloes non-randomly within the haloes.}

Although it is encouraging that it is possible to find simple models in which
it is possible to understand the observed correlation properties of galaxies, 
there are other longstanding puzzles concerning the galaxy distribution.
Arguably the chief of these concerns the dynamical properties
of galaxies, in particular the pairwise peculiar velocity dispersion. 
This statistic has been the subject of debate, and preferred
values have crept up in recent years, to perhaps 450 or $500 \kms$
at projected separations around 1~Mpc (e.g. Jing, Mo \& B\"orner 1998),
most simple models predict a higher figure.
Clearly, the amplitude of peculiar velocities depends on the
normalization of the fluctuation spectrum; however, if this
is set from the abundance of rich clusters, then Jenkins et al. (1998)
found that reasonable values were predicted for large-scale
streaming velocities, independent of $\Omega$. However,
Jenkins et al. also found a robust prediction for the pairwise peculiar velocity dispersion
around 1~Mpc of about $800\kms$. The observed galaxy velocity
field appears to have a higher `cosmic Mach number' than the
predicted dark-matter distribution.

This difficulty is also solved by the simple bias model
discussed here. Two factors contribute: the variation of
occupation number with mass downweights the contribution of more
massive groups, with larger velocity dispersions.
Also, where one galaxy is centred on a halo, it gains
a peculiar velocity which is that of the centre of mass of
the halo, but does not reflect the internal velocity dispersion
of the halo. 
Given a full $N$-body simulation, it is easy enough to predict what
would be expected for a realistic bias model: one needs to
construct a halo catalogue, calculating the peculiar velocities
and internal velocity dispersions of each halo. Knowing the
occupation number as a function of mass, a montecarlo catalogue
of `galaxies' complete with peculiar velocities can be generated.
As shown in figure 15, the effect of the empirical bias recipe
advocated here is sufficient
to reduce the predicted dispersion into agreement with observation.
The simple model outlined here thus gives a consistent picture,
and it is tempting to believe that it may capture some of the
main features of realistic models for galaxy bias.

\japfig{57}{211}{515}{588}{japfig15.eps}
{The line-of-sight pairwise velocity dispersion for the $\Lambda$CDM model.
The top curve shows the results for all the mass; the lower pair
of curves shows the predicted galaxy results, with and without 
assuming that one galaxy occupies the halo centre (the former case gives
the lowest curve).}

\sec{Conclusions}

It should be clear from these lectures that large-scale structure has 
advanced enormously as a field in the past two decades.
Many of our long-standing ambitions have been realised;
in some cases, much faster than we might have expected.
Of course, solutions for old problems generate new difficulties.
We now have good measurements of the clustering
spectrum and its evolution, and it is arguable that the
discussion of section 7.4 captures the main features of the
placement of galaxies with respect to the mass.
However, a fairly safe bet is that one of the
major results from new large surveys such as 2dF and
Sloan will be a heightened appreciation of the subtleties
of this problem.

Nevertheless, we should not be depressed if problems
remain. Observationally, we are moving from an era of
20\% -- 50\% accuracy in measures of large-scale
structure to a future of pinpoint precision. This
maturing of the subject will demand more careful
analysis and rejection of some of our existing tools
and habits of working. The prize for rising to this
challenge will be the ability to claim a real understanding
of the development of structure in the universe. We are
not there yet, but there is a real prospect that the next
5--10 years may see this remarkable goal achieved.

\section*{Acknowledgements}

I thank my colleagues in the 2dF Galaxy Redshift Survey
for permission to reproduce our joint results in section
5.3, and Robert Smith for the joint work reported
in section 7.4.

\section*{References}

\japref Ballinger W.E., Peacock J.A., Heavens A.F., 1996, MNRAS, 282, 877
\japref Baugh C.M., Efstathiou G., 1993, MNRAS, 265, 145
\japref Baugh C.M., Efstathiou G., 1994, MNRAS, 267, 323
\japref Benson A.J., Cole S., Frenk C.S., Baugh C.M., Lacey C.G., 1999, astro-ph/9903343
\japref Bond J.R., Cole S., Efstathiou G., Kaiser N., 1991, \apj, 379, 440
\japref Bond J.R., 1995, \prl, 74, 4369
\japref Bunn E.F., 1995, PhD thesis, Univ. of California, Berkeley
\japref Coles P., {1993}, {MNRAS}, {262}, {1065}
\japref Colless M., 1999, Phil. Trans. R. Soc. Lond. A, 357, 105
\japref Davis M., Peebles P.J.E., 1983, \apj, 267, 465
\japref Dekel A., Lahav O., 1999, \apj, 520, 24
\japref Efstathiou G., Sutherland W., Maddox S.J., 1990,  Nature, 348, 705
\japref Eisenstein D.J., Hu W., 1998, ApJ, 496, 605
\japref Eisenstein D.J., Zaldarriaga M., 1999, astro-ph/9912149
\japref Eke V.R., Cole S., Frenk C.S., 1996, \mn, 282, 263
\japref Feldman H.A., Kaiser N., Peacock J.A., 1994, \apj, 426, 23
\japref Folkes S., et al., 1999, \mn, 308, 459
\japref Gazta\~naga E., Baugh C.M. 1998, \mn, 294, 229
\japref Ghigna S., Moore B., Governato F., Lake G., Quinn T., Stadel J., 1998, \mn, 300, 146
\japref Goldberg D.M., Strauss M.A., 1998, ApJ, 495, 29
\japref Hamilton A.J.S., Kumar P., Lu E.,  Matthews A., 1991, \apj, 374, L1 
\japref Hamilton A.J.S., 1992, \apj, 385, L5
\japref Hamilton A.J.S., 1997a, astro-ph/9708102
\japref Hamilton A.J.S., 1997b, \mn, 289, 285
\japref Hamilton A.J.S., 1997c, \mn, 289, 295
\japref Huchra J.P., Geller M.J., de Lapparant V., Corwin H.G., 1990, \apjs, 72, 433
\japref Jain B., Mo H.J., White S.D.M., 1995, \mn, 276, L25 
\japref Jenkins A., Frenk C.S., Pearce F.R., Thomas P.A., Colberg J.M., White S.D.M., Couchman H.M.P., Peacock J.A., Efstathiou G., Nelson A.H., 1998, ApJ, 499, 20
\japref Jing Y.P., Mo H.J., B\"orner G., 1998, ApJ, 494, 1
\japref Kaiser N., 1987, MNRAS, 227, 1
\japref Klypin A., Primack J., Holtzman J., 1996, \apj, 466, 13
\japref Klypin A., Gottloeber S., Kravtsov A.V., Khokhlov A.M., 1999, 516, 530
\japref Ma C.-P., 1999, \apj, 510, 32
\japref Maddox S. Efstathiou G., Sutherland W.J., 1996, MNRAS, 283, 1227
\japref Mann R.G., Peacock J.A., Heavens A.F., {1998}, MNRAS, 293, 209
\japref Margon B., 1999, Phil. Trans. R. Soc. Lond. A, 357, 93
\japref Matsubara T., Szalay A.S., Landy S.D., 1999, astro-ph/9911151
\japref Meiksin A.A., White M., Peacock J.A., MNRAS, 1999, 304, 851
\japref Moore B., Frenk C.S., White S.D.M., 1993, \mn, 261, 827
\japref Moore B., Quinn T., Governato F., Stadel J., Lake G., 1999, astro-ph/9903164
\japref Navarro J.F., Frenk C.S., White S.D.M., 1996, ApJ, 462, 563
\japref Neyman, Scott \& Shane 1953, ApJ, 117, 92
\japref Padmanabhan N., Tegmark M.,  Hamilton A.J.S., 1999, astro-ph/9911421 
\japref Peacock J.A., Dodds S.J., 1994, MNRAS, 267, 1020
\japref Peacock J.A., Dodds S.J., 1996, MNRAS, 280, L19
\japref Peacock J.A., 1997, \mn, 284, 885
\japref Pearce F.R., et al., 1999, astro-ph/9905160
\japref Peebles P.J.E., 1974, A\&A, 32, 197
\japref Pen, W.-L., 1998, ApJ, 504, 601
\japref Saunders W., et al., 1999, astro-ph/9909191 
\japref Shectman S.A., Landy S.D., Oemler A., Tucker D.L., Lin H., Kirshner R.P., Schechter P.L., 1996, \apj, 470, 172
\japref Sheth R.K., Tormen G., 1999, astro-ph/9901122
\japref Strauss M.A., Willick J.A., 1995, Physics Reports, 261, 271
\japref Sugiyama N., {1995}, {\apjs}, {100}, {281}
\japref Tegmark M., 1996, \mn, 280, 299
\japref Tegmark M., Taylor A.N., Heavens A.F., 1997, \apj, 480, 22
\japref Tegmark M., Hamilton A.J.S., Strauss M.A., Vogeley M.S., Szalay A.S., 1998, \apj, 499, 555
\japref Viana P.T., Liddle A.R., 1996, MNRAS, 281, 323
\japref Vogeley M.S., Szalay, A.S., 1996, \apj, 465, 34
\japref White S.D.M., Rees M., 1978, \mnras, 183, 341
\japref White S.D.M., Efstathiou G., Frenk C.S., 1993, {\mn}, {262}, 1023
\japref Yano T., Gouda N., 1999, astro-ph/9906375

\end{document}